\documentclass[a4paper]{article}
\usepackage{multirow}
\usepackage{amssymb,CJK, indentfirst,balance,appendix}
\usepackage{float}
\usepackage{ISCSLP2022}
\usepackage[multiple]{footmisc}

\title{The ISCSLP 2022 Intelligent Cockpit Speech Recognition Challenge (ICSRC): Dataset, Tracks, Baseline and Results}
\name{Ao Zhang$^1$, Fan Yu$^1$, Kaixun Huang$^1$, Lei Xie$^{1*}$\thanks{* Corresponding author.}, Longbiao Wang$^2$, Eng Siong Chng$^3$, Hui Bu$^4$, \\ Binbin Zhang$^5$, Wei Chen$^6$, Xin Xu$^4$}
\address{
  $^1$Audio, Speech and Language Processing Group (ASLP@NPU), School of Computer Science, Northwestern Polytechnical University, Xi’an, China\\
  $^2$Tianjin Key Laboratory of Cognitive Computing and Application,College of Intelligence and Computing, Tianjin University, Tianjin, China\\
  $^3$School of Computer Science and Engineering, Nanyang Technological University, Singapore\\
  $^4$Beijing Shell Shell Technology Co. Ltd, Beijing, China\\
  $^5$WeNet Open Source Community 
  $^6$Li Auto Inc., Beijing, China}
\email{zhang1998@mail.nwpu.edu.cn, lxie@nwpu.edu.cn}

\begin{document}
 
\maketitle
\begin{abstract}
This paper summarizes the outcomes from the ISCSLP 2022 Intelligent Cockpit Speech Recognition Challenge (ICSRC). We first address the necessity of the challenge and then introduce the associated dataset collected from a new-energy vehicle (NEV) covering a variety of cockpit acoustic conditions and linguistic contents. We then describe the track arrangement and the baseline system. Specifically, we set up two tracks in terms of allowed model/system size to investigate resource-constrained and -unconstrained setups, targeting to vehicle embedded as well as cloud ASR systems respectively. Finally we summarize the challenge results and provide the major observations from the submitted systems.

\end{abstract}
\noindent\textbf{Index Terms}: Automatic speech recognition, intelligent cockpit, in-vehicle speech recognition

\section{Introduction}



As cars become an indispensable part of human daily life, a secure and comfortable driving environment is more and more desirable.
The touch-based interaction in the traditional cockpit is easy to distract the driver's attention, leading to inefficient operations and potential security risks.
Thus the concept of the \textit{intelligent cockpit} is gradually on the rise~\cite{li2021deep,bhat2018tools}, which aims to achieve a seamless driving and cabin experience for drivers and passengers by integrating multimodal intelligent interactions, like speech, gestures, body, etc., with different functions, like commands recognition, entertainment, navigation, etc~\cite{li2021visual}.
As a natural user interface (NUI), a robust automatic speech recognition (ASR) system is crucial~\cite{li2021spontaneous,zhou2019hidden}. In recent years, ASR systems, particularly those end-to-end (E2E) neural approaches, such as connectionist temporal classification (CTC)~\cite{graves2006connectionist,graves2014towards}, recurrent neural network transducer (RNN-T)~\cite{graves2012sequence}, attention-based encoder-decoder (AED)~\cite{chan2016listen,bahdanau2016end,dong2018speech}, have been gradually employed in many applications such as voice assistant, video captioning and meeting transcription. Although speech interface has been recently adopted in many cars, there are still some significant challenges for ASR in the driving scenario.

First of all, the acoustic environment of the cockpit is complex. Since the cockpit is a closed and irregular space, it has special room impulse response (RIR), resulting in special reverberation conditions.  In addition, there are various kinds of noise during driving from both inside and outside, such as wind, engine, wheel, background music and interfering speaker, etc. Although recent advances on speech front-end processing and robust speech recognition can be leveraged, the special acoustic conditions mentioned above should be particularly considered. We notice that researchers have started to explore different techniques to reduce noise in the vehicle for building a robust speech recognition for intelligent cockpit~\cite{prodeus2015performance,zhang2003csa}. Separating different speakers and realizing multi-talker ASR in a vehicle are also desired to truly facilitate the needs of both driver and passenger(s).

Another challenge lies in the linguistic content to be recognized by a cockpit ASR system. Specifically, the main content of intelligent cockpit speech interaction is the user's commands, which includes controlling the air conditioner, playing songs, navigating, etc. These commands may involve a large number of named entities such as contacts, singer names, and points of interest (POI). Recognizing these phrases in the cockpit scene remains to be improved because most named entities are rare in training data. For instance, the commands for navigation involve lots of place names that are hard to recognize while it's important to recognize them correctly for better user experience. This problem is essentially a domain mismatch problem in linguistic aspects. To solve this problem, previous works have explored the methods of fusing an external language model into the decoding process to inject static contextual knowledge~\cite{toshniwal2018comparison,kannan2018analysis,zhao2019shallow}. In contrast, injecting contextual knowledge dynamically is also effective, such as attention-based deep context~\cite{pundak2018deep,jain2020contextual} and trie-based deep biasing~\cite{le2021contextualized,le2021deep}. Besides the language aspects, more challengingly, detecting to whom (voice assistant or passengers) the speaker is talking is also essential in a full-duplex speech interaction system. Moreover, the use of multi-modal integration, such as audio-visual information, is definitely beneficial to smart cockpit systems including ASR~\cite{ivanko2022davis,kashevnik2021multimodal}.

Data sparsity is another obstacle to the related research in intelligent cockpit scenario. This stands in stark contrast to the recent proliferating of open-source speech data covering multiple domains and languages\footnote{https://www.openslr.org}. This is probably because the speech collection process itself in cabin environments is difficult and covering various acoustic conditions during recording is even more challenging~\cite{rus}.


To trigger speech recognition and related research in the intelligent cockpit scenario, we launch the Intelligent Cockpit Speech Recognition Challenge (ICSRC)\footnote{https://iscslp2022-icsrc.org} in ISCSLP2022. In this challenge, a 20-hour Mandarin speech dataset recorded in a new-energy vehicle (NEV) is available to participants. This dataset covers a variety of acoustic conditions and linguistic contents. We particularly set up two tracks in terms of allowed model/system size to investigate resource-constrained and -unconstrained setups, targeting to vehicle embedded as well as cloud ASR systems respectively. This paper introduces the details of the dataset, arrangement on challenge tracks, baseline system and summary on the challenge results. We hope the open data and the challenge will serve as a benchmark and common test-bed for in-vehicle speech recognition.


\section{Data}  

The released dataset of the challenge contains 20 hours of speech data in total, which is divided into 10 hours for evaluation (\textit{Eval}) in the model training phase and another 11 hours as \textit{Test} set for challenge scoring and ranking.
Both Eval and Test sets have 50 non-overlapped speakers with balanced gender coverage. The Eval set is released to the participants at the beginning of the challenge and can be used for finetuning model, while the Test set is released at the final challenge scoring stage. For the \textit{Training} set, participants can only use the constrained data allowed by the challenge, which are open source corpora selected from openslr.org.
More details of the released data are shown in Table~\ref{tab:tongji}.

\begin{table}[!htb]
\centering
\caption{Details of the released Eval and Test data.}
\vspace{0pt}
\begin{tabular}{lllc}
\toprule
Dataset & Uttrance & Duration (h) & Format                                                                                    \\ \midrule
Eval    & 8,780     & 10.01       & \multirow{2}{*}{\begin{tabular}[c]{@{}c@{}}16kHz, 16bit\end{tabular}} \\
Test    & 9,509     & 11.06       &                                                                                            \\ 
\bottomrule
\end{tabular}
\label{tab:tongji}
\end{table}

\begin{figure}[h]
	\includegraphics[width =0.95\linewidth]{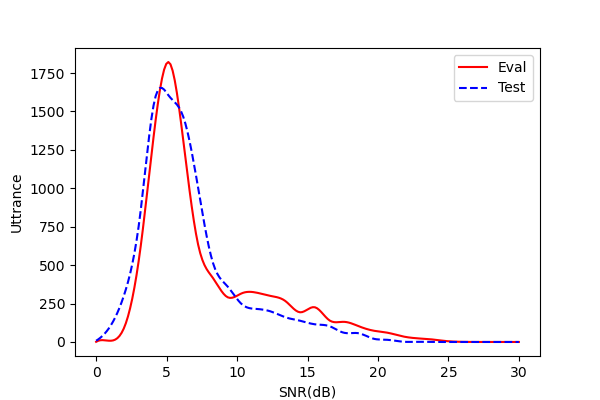}
	\caption{
	SNR statistics of the released Eval and Test data.
	}
	\label{tab:snr}
\vspace{-16pt}
\end{figure}

The data is collected in the NEV with a Hi-Fi microphone placed on the display screen of the car.
During recording, a speaker sits on the front passenger seat. The distance between the microphone and the speaker is around 0.5 m. All speakers are native Chinese speaking Mandarin without strong accents. During driving, the driver may drives on different road conditions (city streets and highways) with various driving speed, open and close windows and play music, which cover various acoustic conditions. 
The data distribution in terms of signal-to-noise ratios (SNR) is shown in Figure
~\ref{tab:snr}. We can see that the SNR covers a wide range and a large part of the data is collected with strong noise. According to the linguistic content, the dataset can be categorized into five categories listed as follows and the statistics are further detailed in Table~\ref{tab:info}.

\begin{CJK*}{UTF8}{gbsn}
\begin{itemize}
    \item \textbf{Air Conditioner}: includes commands to control the air conditioner, like ``调高温度 (turn up the temperature)'';
    \item \textbf{Phone Call}: includes commands to make a phone call, like ``呼叫张奥 (call Ao Zhang)'';
    \item \textbf{Music}: includes commands to control media player, like ``播放周杰伦的歌 (play Jay Zhou's song)'';
    \item \textbf{Point of Interest (POI)}: includes commands to navigation, like ``导航到博物馆 (route to the museum)'';
    \item \textbf{Others}: instead of commands, like reading news, conversations between passengers.
\end{itemize}
\end{CJK*}

\begin{table}[!htb]
\centering
\tabcolsep=0.1cm
\caption{The percentage (\%) of each category for the dataset.}
\vspace{0pt}
\begin{tabular}{cccccc} 
\toprule
{Category} & Air Cond.           & Phone Call  & Music   & POI  & Others             \\ \midrule
Eval       & 15                          & 10          & 15      &15    & 45        \\
Test       & 18                          & 12          & 13      &16    & 41        \\

\bottomrule
\end{tabular}
\label{tab:info}
\end{table}


\vspace{-0.1cm}

\section{Tracks, Evaluation and Baselines}
In the challenge, we design two tracks for participants to select, which evaluated in typical error metric of ASR. We also release a baseline system on github for quick start.
\subsection{Track setting}

In real applications, the demand for intelligent cockpit speech recognition system is twofold. With better Internet connection, one can run on the cloud side with larger model and higher performance. Another is desired to run locally on the vehicle with smaller footprint, lower latency as well as better privacy. Therefore, we simply set up two tracks in the challenge in terms of the limitation on the model size:
\begin{itemize}
\item Track I (Limited model size): The number of model parameters cannot exceed 15M and the finite state transducer (FST) file of the system should be less than 25M if participants build a system with FST.
\item Track II (Unlimited model size): The number of model parameters is not limited.
\end{itemize}

\subsection{Evaluation Metrics}
In the challenge, the performance of an ASR system is measured by the typical \textit{Character Error Rate} (CER). The CER indicates the percentage of Mandarin characters that are incorrectly predicted. Specifically for a given hypothesis output, it measures the ratio of character errors, including insertions (Ins), substitutions (Subs) and deletions (Del), to the total characters spoken in an utterance under the dynamic programming algorithm. According to the challenge rule, if the CERs of two systems on the Test set are coincidentally the same, the one with lower model size and computation complexity will be judged as the superior one. 

\subsection{Baseline}
We release \textit{baseline} recipes\footnote{https://github.com/smallsmartao/ICSRC\_2022\_baseline} based on the WeNet~\cite{wenet} end-to-end speech recognition toolkit for quick start and reproducible research. Specifically, our baseline ASR system follows the standard configuration of the state-of-the-art Conformer structure -- a 12-block encoder and a 6-block decoder, where the dimension of attention and feed-forward layer is set to 256 and 2048 respectively. In all attention sub-layers, the head number of the multi-head attention is set to 4. As for the acoustic feartures, we use 80-dimensional mel-filterbanks (FBank) with a 25ms frame length and a 10ms shift. The whole network is trained using the open-source AISHELL-1~\cite{aishell1} corpus for 240 epochs and warm-up is used for the first 25,000 iterations. The training objective is a logarithmic linear combination of the CTC and attention objectives.
For inference, we adopt the attention-rescoring decoding strategy as default. We also apply SpecAugment~\cite{specaug} for all data during model training.



\section{Results and Discussion}
Twenty five teams finally submitted their results to the challenge, and the CER for the top 4 teams of each track is summarized in Table~\ref{tab:result}.
Observing the detailed performance by subsets of different category, we can see that the phone call and POI subsets, which involve contacts and location addresses, are still challenging for a general speech recognizer with the current limited training data.
Specifically, there are clear performance gaps between the air conditioner subset which is mostly composed of fixed control commands and the other subsets which are much richer in linguistic content.

The major techniques used by the top ranking teams, including network structure, data augmentation strategy, front-end processing, language modeling and training strategy, are summarized in Table~\ref{tab:network} and discussed in the the following subsections accordingly.

We notice that the winner goes to team \textit{Toong} (T044) in both tracks with the lowest CER of 10.66\% and 8.94\% respectively, surpassing the official baseline (53.57\%/47.98\%) with a large margin. The good performances are obtained by
a CTC model for Track I and a CTC/AED hybrid model for Track II respectively. Particularly, the superior performance is attributed largely to a 6-gram language model with iterative interpolation and pruning strategy. Moreover, system fusion is also
beneficial as reported by the winner team.

\begin{table}[!htb]
\centering

\caption{CERs on test set for top ranking teams and baselines (BSL)}
\centering
\setlength{\tabcolsep}{3.5pt}

\begin{tabular}{cccccccc}
\toprule
Track              & \begin{tabular}[c]{@{}c@{}}Team \\ Code\end{tabular} &  \begin{tabular}[c]{@{}c@{}}Air \\ Cond.\end{tabular} & \begin{tabular}[c]{@{}c@{}}Phone \\ Call\end{tabular} & Music & POI   & Others  &All  \\ \midrule
\multirow{5}{*}{1} 
                   & T044                                                  & 1.86                                                       & 14.26      & 9.74  & 21.23 & 10.93  &10.66 \\
                   & T002                                                  & 3.45                                                       & 21.06      & 10.86 & 27.18 & 11.36  &12.67\\
                   & T009                                                  & 2.63                                                       & 15.78      & 12.57 & 24.31 & 14.54  &13.39\\
                   & T013                                                  & 1.42                                                       & 15.75      & 13.49 & 25.66 & 14.89  &13.57 \\ 
                   & BSL                                              & 45.30                                                       & 61.98      & 62.03  & 67.48 & 50.37   &53.57 \\\midrule
\multirow{5}{*}{2} 
                   & T044                                                  & 1.54                                                       & 13.78      & 7.22  & 19.11 & 8.66   &8.94 \\
                   & T009                                                  & 1.35                                                       & 12.18      & 8.78  & 17.34 & 11.03 &9.86 \\
                   & T039                                                  & 1.17                                                       & 13.96      & 9.79  & 21.56 & 10.25 &10.20  \\
                   & T002                                                  & 3.08                                                       & 19.62      & 8.89  & 24.54 & 7.88 &10.21 \\ 
                   & BSL                                              & 40.86                                                       & 57.03      & 57.20  & 61.82 & 44.12  &47.98  \\ 
                   \bottomrule
\end{tabular}
\label{tab:result}
\end{table}

\subsection{Network Structure}
The network structures used by the top 4 teams of both tracks are summarized in Table~\ref{tab:network}.
We can see that most teams adopt the transformer-based E2E neural model, based on an encoder-decoder structure with an attention mechanism.
Specifically, the Conformer architecture~\cite{conformer} is mostly employed as the encoder.
Moreover, team T039~\cite{t039} employs the recently proposed Squeezefomer~\cite{squeeze} as the encoder, which re-examines the design choices for both the macro and micro-architecture of Conformer.
Considering the model size limitation in Track I, most teams choose to adopt fewer network layers or smaller hidden size.
Note that the winner team T044 even abandons the decoder and only keeps the encoder with CTC as the acoustic model.
In Track II, without the limitation on model size, the winner team T044 employs an extremely large model with 574M parameters, which exceeds the model size of other teams obviously, to bring great modeling ability. 
Team T009~\cite{t009} adopts self-conditioned folded encoder~\cite{fold} and inter-CTC loss~\cite{interctc} to perform iterative refinement implicitly by stacking multiple shared encoder layers with fewer parameters, which leads to 6.59\% relative CER reduction on the Test set (12.89\% to 12.04\%). 
Team T039 adds a predictor and RNN-T loss to the transformer-based model and the integrated model, which leads to 0.48\% absolute CER reduction compared with the original transformer-based model. Besides, ROVER~\cite{fiscus1997post} is also adopted by this team to fuse the multiple outputs from different ASR models, which brings 14.93\% relative CER reduction (from 11.90\% to 10.2\%) on the Test set.


\begin{table*}[]
\caption{Major techniques from top performing teams. Here En and De stand for encoder and decoder respectively.}
\renewcommand\arraystretch{1.05}
\begin{tabular}{ccccccccc}
\toprule
Track              & Rank & Team & \multicolumn{1}{c}{Data Aug.}                                          & Network                                                                    & Loss                                                       & Model Size (M) & LM \& TLG Size                                                         & CER (\%) \\ \midrule
\multirow{4}{*}{1} & 1    & T044      & \begin{tabular}[c]{@{}c@{}}Spd,Ptch,Ns,\\ Rvb,SpcAug\end{tabular}     & En: 5 Cfmr Blk                                                            & CTC                                                        & 12.34        & \begin{tabular}[c]{@{}c@{}}6-gram\\ 25M\end{tabular}       & 10.66   \\ \cline{2-9} 
                   & 2    & T002      & \begin{tabular}[c]{@{}c@{}}Spd,Ptch,Ns,\\ Rvb,SpcAug\end{tabular}     & \begin{tabular}[c]{@{}c@{}}En: 5 Cfmr Blk\\ De: 4 Tfmr Blk\end{tabular}   & CTC Att                                                    & 14.6         & \begin{tabular}[c]{@{}c@{}}3-gram\\ 24M\end{tabular}       & 12.67   \\ \cline{2-9} 
                   & 3    & T009      & \begin{tabular}[c]{@{}c@{}}Spd,Ptch,Ns,\\ Rvb,SpcAug,TTS\end{tabular} & \begin{tabular}[c]{@{}c@{}}En: 12 Cfmr Blk\\ De: 2 Tfmr Blk\end{tabular}  & \begin{tabular}[c]{@{}c@{}}CTC Att\\ interCTC\end{tabular} & 14.82        & \begin{tabular}[c]{@{}c@{}}4-gram\\ 24M\end{tabular}       & 13.39   \\ \cline{2-9} 
                   & 4    & T013      & \begin{tabular}[c]{@{}c@{}}Spd,Ptch,Ns,\\ Rvb,SpcAug\end{tabular}     & \begin{tabular}[c]{@{}c@{}}En: 6 Cfmr Blk\\ De: 2 Tfmr Blk\end{tabular}   & CTC Att                                                    & 14.87        & -                                                          & 13.57   \\ \hline
\multirow{4}{*}{2} & 1    & T044      & \begin{tabular}[c]{@{}c@{}}Spd,Ptch,Ns,\\ Rvb,SpcAug\end{tabular}     & \begin{tabular}[c]{@{}c@{}}En: 12 Cfmr Blk\\ De: 12 Tfmr Blk\end{tabular} & CTC Att                                                    & 574          & \begin{tabular}[c]{@{}c@{}}6-gram\\ 5.7G\end{tabular}      & 8.94    \\ \cline{2-9} 
                   & 2    & T009      & \begin{tabular}[c]{@{}c@{}}Spd,Ptch,Ns,\\ Rvb,SpcAug,TTS\end{tabular} & \begin{tabular}[c]{@{}c@{}}En: 15 Cfmr Blk\\ De: 3 Tfmr Blk\end{tabular}  & \begin{tabular}[c]{@{}c@{}}CTC Att\\ interCTC\end{tabular} & 148          & \begin{tabular}[c]{@{}c@{}}4-gram+Tfmr\\ 410M\end{tabular} & 9.86    \\ \cline{2-9} 
                   & 3    & T039      & \begin{tabular}[c]{@{}c@{}}Spd,Ns,Rvb,\\ SpcAug,TTS\end{tabular}      & \begin{tabular}[c]{@{}c@{}}En: 12 Sfmr Blk\\ De: 6 Tfmr Blk\end{tabular}  & \begin{tabular}[c]{@{}c@{}}CTC Att\\ RNN-T\end{tabular}    & 149          & LSTM                                                       & 10.20   \\ \cline{2-9} 
                   & 4    & T002      & \begin{tabular}[c]{@{}c@{}}Spd,Ptch,Ns,\\ Rvb,SpcAug\end{tabular}     & \begin{tabular}[c]{@{}c@{}}En: 15 Cfmr Blk\\ De: 6 Tfmr Blk\end{tabular}  & CTC Att                                                    & 46           & \begin{tabular}[c]{@{}c@{}}3-gram\\ 360M\end{tabular}      & 10.21   \\ \bottomrule
\end{tabular}\vspace{3pt}
\\\footnotesize{Note: Spd--Speed perturb.; Ptch--Pitch shift; Ns--Noise augment.; Rvb--Reverb simu.; Cfmr--Conformer; Tfmer-Transformer; Sfmer--Suqeezeformer~\cite{squeeze}}
\label{tab:network}
\vspace{-0.6cm}
\end{table*}


\subsection{Data Augmentation and Speech Enhancement}
Since the acoustic environment of the cockpit is complex, reflected from the released data, data augmentation is adopted by most teams to achieve better data coverage.
 Noise and reverberation augmentation is widely used. 
Moreover, most teams adopt speed perturbation and spec-augment to further augment the training samples, which are the mainstream tricks with stable performance improvement.
Pitch shift is also an effective data augmentation method reported by several teams, which changes the pitch of a speech signal without altering its duration.
It is worth noticing that team T009 makes particular efforts in data augmentation besides the above typical tricks.
The first special trick is random cutting and splicing a speech utterance. In detail, two pieces of audio of the same speaker are randomly selected from the training set, and then each piece is cut into two splices and the splices from the two pieces are randomly combined as new samples.
Secondly, team T009 uses the provided data to train a TTS model particularly for producing extra speech for model training. With the recent advances in speech generation~\cite{tanxu}, TTS based data augmentation has gradually become a promising solution for ASR in constrained data scenarios~\cite{tts2,tts1}.  

Speech enhancement is conducted by team T039, where a DCCRN~\cite{dccrn} model is adopted to remove noise from the released cabin speech. Team T033 adopts volume normalization particularly on audio signal before model training.


\subsection{Language Modeling}
Considering the linguistic characteristics of cabin speech, language model plays a significant role for a in-vehicle ASR system. Accordingly, most teams use language models based on the provided data. As show in Table~\ref{tab:network}, we summarize the language modeling methods of the top 4 teams in the challenge.

As for language modeling, it is obvious that fusing an
n-gram language model with the E2E acoustic model works well, and it brings up to 2\% relative improvement on the Test set, reported from several submitted system descriptions.
Moreover, with the premise that a larger N-gram can obtain richer contextual information, the winning team T044 employs a 6-gram LM for both tracks, which leads to substantial performance gain according to its system description. For generating a high quality N-gram LM, they adopt a 3-stage based iterative strategy. Firstly, a 6-gram LM is trained on the transcripts of the Training set. To eliminate the domain mismatch between the general Training set and the released Eval set, they select the unincluded N-gram phrases from the Eval set based on the well-trained 6-gram LM.
Secondly, they randomly replace the original phrases in the Training transcripts with the selected unincluded phrases according to their parts-of-speech and simulate more in-domain texts.
Finally, a new LM is trained based on the simulated text and is composed with the well-trained LM.
They repeat aforementioned stages multiple times until there are no unincluded phrased in the Eval set. 
In Track I, the FST file of a system should be less than 25M, which means that the participants need to particularly prune the graph as much as possible and control the amount of data used to build the language model. 
Thus, data filtering on the training data is used by Team T042, which produces a language model better matched to the intelligent scene.

Different from Track I, Track II does not has model size limitation. Thus several teams adopt a neural network language model (NNLM) with large model size. 
NNLM with different structures, including LSTM and Transformer, are adopted by team T039 and T035 for shallow fusion.
Moreover, team T009 fuses scores from an n-gram model and a transformer language model in the decoding process.
As reported by several teams, the contribution from data filtering is not salient for NNLM, in contrast to the greater contribution to the n-gram language model.

In summary, for both tracks, language model is proved to be an effective method to bias the recognition results towards the desired domain. However, from Table~\ref{tab:result}, we can find that the commands involving named entities are still challenging for a general speech recognizer trained with the current training data. More effective domain transfer or biasing method should be considered in the future.

\subsection{Training Strategy}
In order to better explore the potential of the E2E models, several teams study some special training strategies. Particularly, team T002 employs curriculum learning strategy~\cite{braun2017curriculum} to the training data ordered by noisy conditions in terms of SNR.
It turns out that the ASR model can achieve better performance when provided with a better organized training set, for example, the one here consisting of training samples that exhibit an decreasing SNR level. In other words, it is beneficial to learn a model from easy to hard samples.
Inspired by knowledge distillation~\cite{hinton2015distilling}, team T009 considers the model in Track II as the teacher model and learns a student model for Track I accordingly.
Specifically, the Track II model (teacher model with more parameters) provides soft labels of current utterance for the Track I model (student model) training, where Kullback Leibler (KL) divergence loss is adopted between the output posterior distributions of the two models to transfer the generalization ability.
This method narrows the gap between the small and large models in the two tracks, leading to 13.87\% relative CER reduction (12.04\% to 10.36\%) for the system in Track I.



\vspace{-0.1cm}
\section{Conclusions}
This paper summarizes the outcomes of the ISCSLP 2022 Intelligent Cockpit Speech Recognition Challenge by introducing the background, tasks, evaluation metric, baseline system as well as the results. We hope that the initiative of the ICSRC challenge and the released data will promote speech recognition and related research in intelligent cockpit scenario. Although recorded in real cabin conditions, the data made available is limited with only 20 hours in this challenge. We plan to record and release more data in the future, hopefully with both multi-channel audio signals and visual information. As mentioned in Section 1, there are many other challenging tasks in in-vehicle speech processing, including speech separation, multi-talker speech recognition and user intent detection, etc. We believe speech interface will play an increasingly important role with the rapid development of NEV industry, triggering more research in speech related area.


\bibliographystyle{IEEEtran}

\bibliography{mybib}


\end{document}